\documentclass[twocolumn]{aastex631}
\usepackage{enumitem}
\usepackage{comment}
\usepackage{graphicx}
\graphicspath{{./figures/}{./}}
\usepackage{hyperref}
\usepackage{soul} 
\usepackage{amsmath}
\usepackage{amssymb}
\usepackage{xspace}
\usepackage{xifthen}

\newcommand{\ie}{i.e.\xspace}
\newcommand{\eg}{e.g.\xspace}

\newcommand{\NEW}[1]{{\textcolor{black}{#1}}}

\mathchardef\mhyphen="2D

\newlength{\dhatheight}

\newcommand{\unit}[1]{\ensuremath{\mathrm{\,#1}}\xspace}
\newcommand{\Myr}{\unit{Myr}}
\newcommand{\Gyr}{\unit{Gyr}}

\newcommand{\km}{\unit{km}}
\newcommand{\kms}{\km \second^{-1}}
\newcommand{\pc}{\unit{pc}}
\newcommand{\kpc}{\unit{kpc}}

\newcommand{\second}{\unit{s}}

\newcommand{\Msun}{\unit{M_\odot}}

\newcommand{\e}{\unit{e^{-}}}

\newcommand{\kpckms}{\kpc \km \second^{-1}}

\newcommand{\secref}[1]{Section~\ref{sec:#1}}

\newcommand{\figref}[1]{Fig.~\ref{fig:#1}}

\newcommand{\bandvar}[2][]{%
  \ifthenelse{\isempty{#1}}{\var{#2}}{\var{#2\_#1}}%
}

\newcommand{\Gaia}{\textit{Gaia}\xspace}

\newcommand{\var}[1]{\ensuremath{\texttt{\MakeUppercase{#1}}}\xspace}

\providecommand\physrep{\ref@jnl{Phys.~Rep.}}%
\providecommand\apjs{\ref@jnl{ApJS}}%
\providecommand{\jcap}{\ref@jnl{JCAP}}%
\usepackage[absolute,overlay]{textpos}

\begin{document}

\shortauthors{Tavangar et al.}

\title{Phase-Spirals Across Galactic Disks II: Using large-scale ``macro-spirals'' in phase-spiral amplitude to derive perturbation times}
% Author list file generated with: mkauthlist 1.2.4+13.gfd3bfb9.dirty 
% mkauthlist DES-2020-0628_author_list.csv authors.tex -f -a order.txt --orcid 

% Author list file generated with: mkauthlist 1.2.4+13.gfd3bfb9.dirty 
% mkauthlist DES-2020-0628_author_list.csv authors.tex -f -a order.txt --orcid 

\author[0000-0001-6584-6144]{Kiyan Tavangar}
\affiliation{Department of Astronomy, Columbia University, New York, NY 10027, USA}

\author[0000-0001-6244-6727]{Kathryn V. Johnston}
\affiliation{Department of Astronomy, Columbia University, New York, NY 10027, USA}

\author[0000-0001-8917-1532]{Jason A. S. Hunt}
\affiliation{School of Mathematics \& Physics, University of Surrey, \\Stag Hill, Guildford, GU2 7XH, UK}

\author[0000-0001-5686-3743]{Axel Widmark}
\affiliation{Department of Astronomy, Columbia University, New York, NY 10027, USA}
\affiliation{Stockholm University and The Oskar Klein Centre for Cosmoparticle Physics, \\
Alba Nova, 10691 Stockholm, Sweden}

\author[0009-0003-2883-7101]{Vandana G. Kaushik}
\affiliation{School of Mathematics \& Physics, University of Surrey, \\Stag Hill, Guildford, GU2 7XH, UK}

\author[0000-0003-0872-7098]{Adrian~M.~Price-Whelan}
\affiliation{Center for Computational Astrophysics, Flatiron Institute, Simons Foundation, 162 Fifth Avenue, New York, NY 10010, USA}

% \author[0000-0003-1517-3935]{Mike Petersen}
% \affiliation{Institute for Astronomy, University of Edinburgh, Royal Observatory, Blackford Hill, Edinburgh EH9 3HJ, UK}

% \author[0000-0003-2660-2889]{Martin Weinberg}
% \affiliation{Department of Astronomy, University of Massachusetts at Amherst, 710 N. Pleasant St., Amherst, MA 01003}

% \author[0000-0002-5861-5687]{Chris Hamilton}
% \affiliation{School of Natural Sciences, Institute for Advanced Study, Princeton, NJ 08540, USA}

% \author{the EXP Collaboration}

\correspondingauthor{Kiyan Tavangar}
\email{k.tavangar@columbia.edu}

\begin{abstract}

Phase-space spirals in the Milky Way disk are a key observable remnant of recent perturbations to the Galaxy.
They provide insight into the disk's potential, dynamical evolution, past interactions, and even substructure.
However, the complex dynamics of phase-spiral formation and evolution have made clear interpretations challenging.
For example, recent work has shown that the ``winding time'' -- measured from how tightly wound a phase-spiral is -- is a biased estimate of the true time since the inciting perturbation due to the complex effects of self-gravity.
In this paper series, we present an alternative approach by looking at \textit{correlations} in phase-spiral morphology across the Galactic disk. 
% In this paper, rather than looking at individual phase-spirals, we use a fully self consistent dynamical simulation to explore how  \textit{correlations} in their properties across the galactic disk evolve following a localized perturbative event. 
Here we show that following a localized perturbative event, the ridgeline connecting the largest amplitude phase-spiral at each radius winds up  into a ``macro-spiral'' at the rate expected for differential rotation.
This means the macro-spiral can 1) be unwound to constrain the origin -- time and location -- of correlated phase-spiral properties and 2) indicate the delay in individual phase-spiral winding as a novel diagnostic of disk dynamical properties.

Applying these ideas to the Milky Way's phase-spirals, we estimate a perturbation time of $\simeq 1$ \Gyr ago, consistent with the penultimate passage of the Sagittarius dwarf galaxy through the disk, and delay times of up to 800 Myr for phase-spirals in the inner disk.
While the current application of this method to \Gaia DR3 data is limited by the available radial velocities, future \Gaia data releases will enable stronger constraints using a much larger area of the Galactic disk.

\end{abstract}


\section{Introduction} \label{sec:intro}

\NEW{Uncovering the history of the Milky Way is an active and evolving area of research. At a broad level, there is evidence that the Milky Way is a barred spiral galaxy that had its last major merger $\gtrsim 6~\Gyr$ ago \citep[\eg][]{Belokurov:2018, Helmi:2018}. However, the details of the Galaxy's more recent, minor mergers are still not well understood. These mergers have affected the dynamics of Galactic stars but the effects are subtle and difficult to spot without large samples and high-resolution observations. In the past decade though, surveys like the \Gaia Mission \citep{gaia:2016} have delivered data with the requisite volume and quality to uncover clear features of disequilibrium (\eg stellar streams, warps, kinematic waves) that may provide the key to the Milky Way's recent past.}

One of the most striking features of disequilibrium in the Milky Way is the vertical phase-space spiral\footnote{These structures are also often referred to as the \Gaia ``snails'' or simply the ``phase-spirals'', which is the terminology we will use in this work.}. % may be a key observable feature for answering some of these questions.
The phase-spiral was initially discovered in \Gaia \citep{gaia:2016} DR2 by \citet{Antoja:2018} by visualizing the density of stars within 0.1 \kpc of the Sun in the $z$--$v_z$ plane; this revealed a spiral structure with an amplitude $\simeq10\%$ of the equilibrium background \citep{Antoja:2023}.
A similar spiral feature exists when the $z$--$v_z$ plane is colored by Galactocentric radial velocity, azimuthal velocity, metallicity, and age \citep[\eg][]{Antoja:2018, Bland-Hawthorn:2019, Frankel:2025}.

The phase-spiral is thought to form following a perturbation to the vertical structure of the disk.
This initially creates a $z$--$v_z$ dipole (bending mode) or quadrupole (breathing mode) depending on the nature of the perturbation \citep{Hunt:2022, Banik:2022, Banik:2023}.
Over time, as stars continue to oscillate vertically with different frequencies, this initial distribution will evolve into a one-armed or two-armed spiral structure (for a  detailed explanation of this process, see the introduction of \citet{Tavangar:2026}, henceforth referred to as Paper I).

A present-day phase-spiral in a given region of the disk therefore encodes information about its origin.
The number of spiral arms indicates whether the perturbation was asymmetric or symmetric in $z$.
Its amplitude (\ie contrast with the equilibrium background) tells us about the force the perturbing object imparted on the disk region and potentially the object's mass.
Its pitch angle (or windedness) depends on both the time of the perturbation and the vertical potential of the disk in that region.

Despite this seemingly straightforward mapping from a phase-spiral to its origin, the cause of the Milky Way's phase-spirals remains unknown.
The most prominent and widely explored hypothesis is that the phase-spirals were caused by the Sagittarius dwarf galaxy (Sgr), a satellite galaxy currently merging with the Milky Way, passing through the disk \citep[\eg][]{Antoja:2018, Binney&Schonrich:2018, Darling&Widrow:2019a, Laporte:2019, Bland-Hawthorn:2019, Hunt:2021, Bennett&Bovy:2021, Gandhi:2022, Bennett:2022, Darragh-Ford:2023}.
Other theories include the buckling of the Galactic bar \citep{Khoperskov:2019}, spiral arms \citep{Faure:2014, Hunt:2022, Li:2023}, Sgr's dynamical friction wake \citep{Grand:2023}, large- and small-scale kicks from substructure in the global potential \citep{Tremaine:2023, Gilman:2025}, and misaligned gas accretion \citep{Wang:2026}.
While each of these ideas can individually cause phase-spirals, none of them can recreate the observations by themselves \citep{Faure:2014, Quillen:2018, Bennett&Bovy:2021}.

For the most part, these origin theories have been tested against the Solar Neighborhood phase-spiral.
However, since the phase-spiral is a local feature, one can split the disk into regions to examine phase-spirals in different parts of the disk.
\citet{Hunt:2022} used \Gaia DR3 data to discover that phase-spiral morphologies vary significantly throughout the Milky Way.
Subsequently, researchers have found large-scale spatial coherence between phase-spiral phase angles \citep{Widmark:2025}, winding times \citep{Frankel:2023, Darragh-Ford:2023, Antoja:2023, Widmark:2025,Hunt&Vasiliev:2025}, and amplitudes \citep{Alinder:2023, Frankel:2023, Frankel:2025}.
In simulations, many studies have built on these observational results by exploring the formation and evolution of phase-spirals across an entire galactic disk \citep[\eg][]{Laporte:2019, Hunt:2021, Alinder:2023, Frankel:2023, Darragh-Ford:2023, Asano&Antoja:2025, Tavangar:2026}, finding similar coherence in phase-spiral properties on large spatial scales.

The correlations between phase-spirals in different parts of the disk are underexplored and may provide an independent way to constrain the phase-spirals' origin, and therefore the Galaxy's recent merger history.
We take a close look at these large-scale correlations between phase-spirals.
In Paper I, we showed that there are significant delays in phase-spiral winding in self-consistent simulations -- compared to a pure phase mixing case -- and that the magnitude of these delays depend strongly on the guiding radii of the stellar orbits.
In this paper, we move on from looking at phase-spirals {\it individually} to examine the {\it population} of phase-spiral amplitudes across a galactic disk. 
Our aim is to understand whether this combination of microscopic (i.e. individual phase-spirals) and macroscopic (i.e. populations of phase-spirals) views might together break the degeneracies found in earlier work and point to more clear methods of constraining the Milky Way's structure and history.

Our paper is organized as follows.
In \secref{methods}, we introduce all the steps in our analysis pipeline, along with the simulation on which we test our methods.
In \secref{data}, we describe the data sample used in this work.
In \secref{results}, we report the results of applying our methods to the simulation and the Milky Way data and place them in the context of recent literature. 
Finally, we summarize and conclude in \secref{conclusion}.

\section{Methods} \label{sec:methods}

\begin{figure*}[th!]
    \centering
    \includegraphics[width=\linewidth]{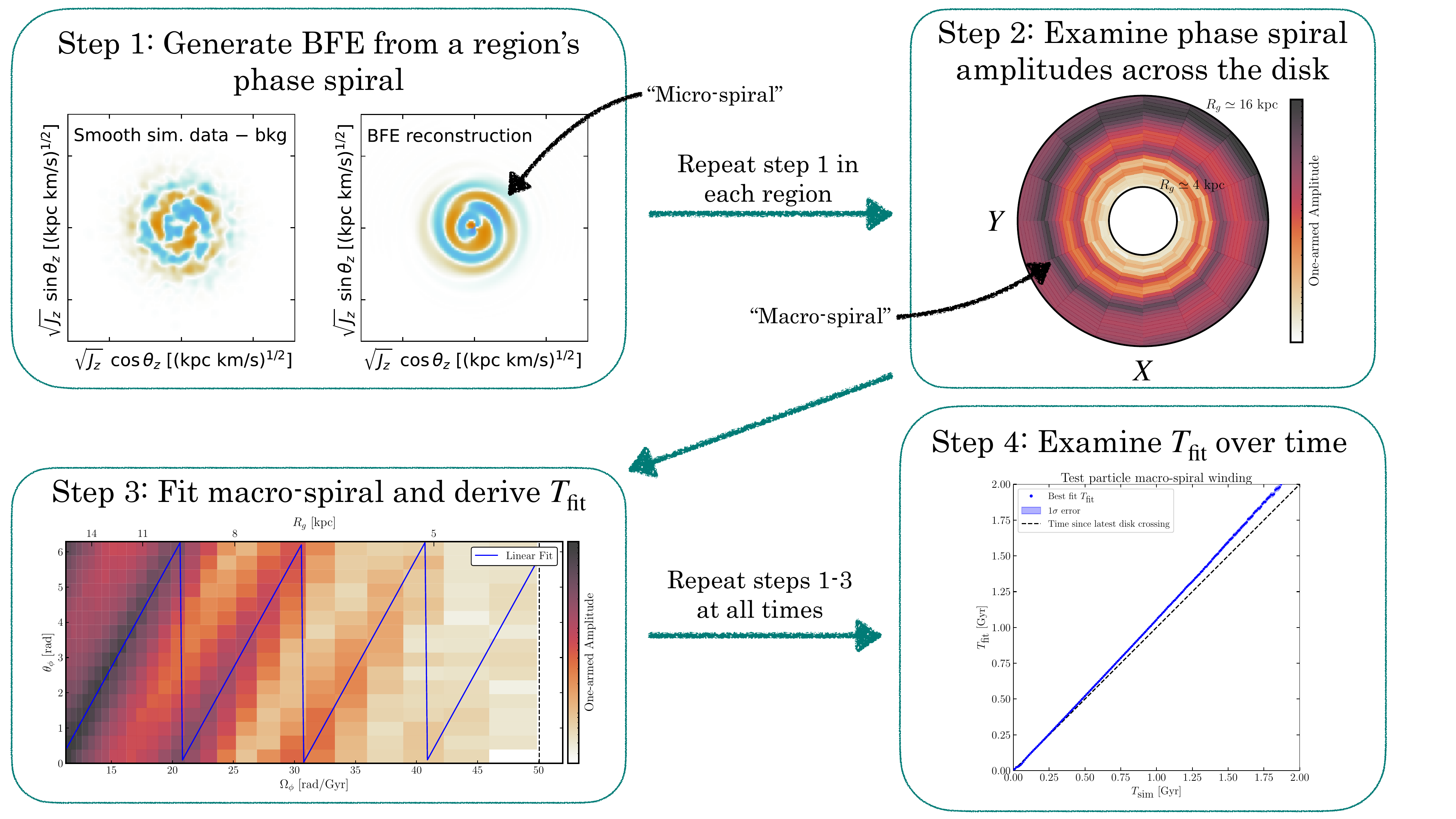}
\caption{The four major steps involved in our analysis applied to an idealized test particle simulation for clear visualization.
\textit{Top Left:} The background-subtracted phase-spiral simulation data in one region (left) and the BFE reconstruction of that phase-spiral from $m=1$ and $m=2$ coefficients (right).
\textit{Top right:} A face-on view of the simulation disk, colored by the one-armed phase-spiral amplitude.
\textit{Bottom Left:} One-armed phase-spiral amplitude in $\theta_\phi - \Omega_\phi$ space along with a fit (in blue) to the high-amplitude ridgeline, the slope of which gives us $T_{\rm fit}$. 
% Step 3 is broken down in further detail in \figref{tfit_pipeline}. 
\textit{Bottom Right:} The recovered macro-spiral winding time from each snapshot of the example test particle simulation.
% \KT{Note to self: using the actual potential and circular velocity was a bit different from using the median as we do for the N-body case in a different context so maybe doing that would take away the imprecision at high simulation times in the Step 4 figure.}
}
\label{fig:analysis_pipeline}
\end{figure*}

This section describes the analysis pipeline applied to the simulated and real data to derive the ``winding time'' from the large-scale spiral pattern apparent in phase-spiral amplitudes.
The steps are summarized in \figref{analysis_pipeline}.%~and~\ref{fig:tfit_pipeline}.
As an overview, we (1) bin particles (stars) to select objects on similar orbits at similar phases (\S \ref{sec:binning}), (2) quantify the amplitude of the vertical phase-spiral in each bin (\ie the ``micro-spirals'') (\S \ref{sec:characterization}), and (3) estimate the global winding time using the pattern of micro-spiral amplitudes (\ie the ``macro-spiral'', \S \ref{sec:tfit_method}).

\subsection{Binning particles on similar orbits} \label{sec:binning}

As in Paper I, we group stars with similar orbits by binning the disk using angle-action coordinates ($\theta_\phi,J_\phi$) rather than physical ones ($x,y,z$).
Stars with the same $\theta_\phi$ and $J_\phi$ have the same azimuthal frequency $\Omega_\phi$ and are at a similar point in their azimuthal orbit, meaning they travel together and have likely experienced similar perturbation histories.
\NEW{By contrast, stars with the same physical coordinates are in the same place at a given time but can have vastly different perturbation histories due to being on very different orbits.
We refer the interested reader to Figure 6 of \citet{Darragh-Ford:2023} for a visual schematic of this idea.}

For reference, in the Milky Way, $J_\phi$ is related to guiding radius $R_g$ via
\begin{equation}
R_g \equiv \frac{J_{\phi}} {v_{\rm circ}(J_\phi)} \simeq 8.2 \: {\rm kpc} \frac{J_\phi}{1900 \: {\rm kpc\; km/s}} \frac{230 \:{\rm km/s}}{v_{\rm circ}}
\end{equation}
where $v_{\rm circ}(J_\phi)$ is the circular velocity of a star with a given $J_\phi$ and the numbers on the right are approximations for the Solar Neighborhood.

\subsection{Characterizing the phase-spirals} \label{sec:characterization}

To generate clean, quantitative descriptions of the phase-spirals in the simulations, we: 
\begin{enumerate}
    \item Convert to vertical angle-action coordinates ($\theta_z, J_z$) to simplify phase-spiral morphology.
    \item Quantify phase-spiral properties using basis function expansions (BFEs).
    \item Extract phase-spiral amplitudes from the constructed BFE.
\end{enumerate}
For the first two steps, see Section 3 of Paper I.
For the third, we use the fact that the amplitudes of $m$-armed spiral features are given by 
\begin{equation}
    C_m = \sqrt{\sum_n A_{nm}^2}
\end{equation}
where $A_{nm}$ are basis coefficients.
From these $C_m$ values, it is straightforward to compare the amplitude of different spiral patterns relative to the monopole ($m=0$), which is assumed to be close to the underlying equilibrium. 

In both our N-body simulation and the Milky Way, the phase-spirals are dominated by one-armed features \citep{Hunt:2021, Hunt:2022}.
We therefore focus on the $m=1$ dipole amplitude by calculating and analyzing the large-scale correlations in $C_1/C_0$.

\subsection{Deriving winding times using the disk's differential rotation} \label{sec:tfit_method}

% \begin{figure*}[th!]
%     \centering
%     \includegraphics[width=\linewidth]{tfit_pipeline.pdf}
% \caption{The details of Step 3 from \figref{analysis_pipeline}. 
% % As in \figref{analysis_pipeline}, the plots shown within this figure are taken from an application of this method to an idealized test particle simulation.
% \textit{Top left:} The one-armed phase-spiral amplitude values for a given snapshot of the test particle simulation. The blue rectangle shows the $\Omega_\phi$ bin used in Step 3b.
% \textit{Top right:} \FIXME{The one-armed phase-spiral amplitude as a function of $\theta_\phi$ for the $\Omega_\phi$ bin highlighted in Step 3a. The solid black curve is the cubic spline fit to these points and the dashed black line shows the peak $\theta_\phi$ of that curve.}
% \textit{Bottom left:} Same background as top left but with the upper and lower $\Omega_\phi$ bounds for our fit shown with dashed green lines.
% \textit{Bottom right:} Same as the bottom left panel of \figref{analysis_pipeline}.}
% \label{fig:tfit_pipeline}
% \end{figure*}

With $C_1/C_0$ for each (micro) phase-spiral across the galactic disk, we can find the macro-spiral winding time. %as shown in \figref{tfit_pipeline}.

The winding time, $T_{\rm fit}$ is defined as the time for the differential rotation of the disk to wind-up a perturbation occurring at azimuth $\theta_{\phi,0}$ into a spiral traced by $\theta_{\phi, \rm max}(J_\phi)$. For a disk with circular speed $v_{\rm circ}(R) \equiv v_{\rm circ}(J_\phi)$:
\begin{equation}
\label{eq:wind_slope}
\theta_{\phi, \rm max}(J_\phi) = T_{\rm fit}\Omega_\phi(J_\phi) + \theta_{\phi, 0}
\end{equation}
where $\Omega_\phi(J_\phi) = v_{\rm circ}(J_\phi)/R_g(J_\phi)$ for guiding radius $R_g$. 
\NEW{This assumes that $\theta_{\phi, 0}$ is independent of $J_\phi$ which is reasonable for any localized perturbation such as a satellite disk crossing.}

Equation \ref{eq:wind_slope} demonstrates that we can derive $T_{\rm fit}$ \NEW{and $\theta_{\phi,0}$} from the slope and intercept of the ridge-line joining the maximum amplitudes in the $\theta_\phi - \Omega_\phi$ plane.
Therefore, we first convert $J_\phi$ into $\Omega_\phi$ by taking the median $\Omega_\phi$ value in each $J_\phi$ annulus and building a simple spline curve to do the mapping.
\NEW{With the resulting $\theta_\phi - \Omega_\phi$ grid of phase spiral amplitudes, we fit $T_{\rm fit}$ and $\theta_{\phi,0}$ values at each timestep using a variant on the Rayleigh test of uniformity \citep{Fisher:1993}.
A Rayleigh test is designed to check whether data points with an angular component to them are uniformly spread around a circle.
In astronomy, this type of analysis is often used to detect periodic signals such as pulsars \citep[\eg][]{Buccheri:1983}.}

\NEW{In our case, we look to take advantage of the fact that at the perturbation time, the highest amplitude phase spirals have azimuthal angles clustered around $\theta_{\phi,0}$ (the azimuthal location of the perturbation).
Simultaneously, consider that because we know the azimuthal frequencies (or circular velocities) of stars in the disk, we can always ``rewind'' a given region to its azimuthal location at a prior time.
This is essentially flipping Equation~\ref{eq:wind_slope} to an equivalent (more general) equation: $\theta_{\rm previous} = \theta_{\rm current} - t_{\rm elapsed} \times \Omega$. 
In our case, $T_{\rm fit}$ at each timestep is the $t_{\rm elapsed}$ value that provides the greatest non-uniformity in the Rayleigh test.}

\NEW{Mathematically, we quantify this non-uniformity by:
\begin{equation}
    U(T_{\rm fit}) = \sum_{j,k} (C_1/C_0)_{j,k} \, \exp\!\big[i\,(\theta_{\phi[j,k]} - T_{\rm fit}\,\Omega_{\phi[j,k]})\big]
\end{equation}
where $j$ and $k$ are indices of the different $\theta_\phi$ and $\Omega_\phi$ bins.
$T_{\rm fit}$ is the value that maximizes $\lvert U(T_{\rm fit})\rvert^2$; at that value, $\theta_{\phi,0} = \arg(U(T_{\rm fit}))$.}

\NEW{We get uncertainties on $T_{\rm fit}$ by bootstrap sampling the phase-spiral amplitudes of each disk region using the errors on each coefficient from the basis function expansion.
We use 100 samples at each timestep.}

\section{Data} \label{sec:data}

We use the \Gaia DR3 data \citep{Gaia:2023, Katz:2023, Recio-Blanco:2023} that has all six dimensions of phase space.
We take the on-sky positions, proper motions, and radial velocities directly from the \Gaia data and use the \citet{Bailer-Jones:2021} \texttt{r\_med\_photogeo} measurements for the distances.
We then remove stars with fractional distance errors (in this case defined by \texttt{r\_hi\_photogeo} - \texttt{r\_lo\_photogeo} / \texttt{r\_med\_photogeo}) greater than 0.1.
We select stars whose cylindrical distance from the Sun ($d_{xy}$) is less than 1 kpc, ensuring we only examine data in regions where dust extinction is less prevalent as was done by \citet{Hunt:2022}\footnote{This selection without a rigorous accounting for extinction can introduce biases with $\theta_r$ and $J_r$ but we do not use those quantities here}.
For the transformation to Galactocentric coordinates, we assume $R_{\odot} = 8.275 \kpc$ \citep{Gravity:2021}, $z_{\odot} = 20.8 \pc$ \citep{Bennett&Bovy:2019}, and a total solar velocity relative to the Galactic center of $\mathbf{v}_{\odot} = (8.4, 251.8, 8.4) \kms$ \citep{Reid&Brunthaler:2020, Gravity:2021}.

Even though we limit our sample to nearby stars, we still apply a simple correction for incompleteness from selection effects to improve the reliability of our BFE characterizations.
This is because our chosen basis set has a particular functional form and an underdensity of stars at low $J_z$ or $\theta_z$ (\ie near the midplane) would create a distribution not naturally fit by our BFEs.
To account for this, we weight stars by the inverse of the probability of their detection.
While a rigorous probabilistic model would instead fold the completeness function directly into a forward model of the population \citep[e.g.,][]{Foreman-Mackey:2014}, here we adopt this simpler approach as a first exploration and defer a proper analysis to a future paper.
We use the \Gaia DR3 RVS selection function from the \texttt{GaiaUnlimited} project\footnote{https://gaia-unlimited.org/} \citep{Rix:2021, Cantat-Gaudin:2023, Castro-Ginard:2023, Cantat-Gaudin:2024} to generate detection probabilities given a star's on-sky position, magnitude, and color.
We remove any stars with undefined weights as well as those with weight $>20$ because we do not want single stars with very large weights to significantly affect the phase space distribution.
This cut eliminates $\simeq 0.1$\% of the original sample, leaving us with 12884718 stars.

To compute the angles and actions $(\boldsymbol{\theta}, \boldsymbol{J})$ of each star we follow the steps from \citet{Hunt:2022}.
Specifically, we assume a four-component Milky Way mass model containing a spherical Herquist nucleus and bulge, an approximately exponential disk and a spherical NFW halo.
We then calculate the angles and actions using the St\"{a}ckel Fudge method \citep{Binney:2012, Sanders:2012} as implemented in \texttt{galpy} \citep{Bovy:2015}.

% Note that we are still using $C_1/C_0$ even in regions where the two armed spiral is visually dominant. This is because a single perturbation is likely to induce the same type of phase-spiral throughout the disk and the BFE formalism allows us to extract a faint one-armed amplitude from our perturbation of interest even when a two-armed feature from another source may be more prominent.

\section{Results} \label{sec:results}

In this section we explore how $T_{\rm fit}$ derived from the macro-spiral across the face of the Galactic disk is related to the known time since interaction in a self-consistent simulation (\S \ref{sec:nbody_results}). We then apply this knowledge to interpret the micro-spiral amplitude oscillations seen in the Milky Way (\S \ref{sec:observations}). 

\subsection{Self-consistent simulation results} \label{sec:nbody_results}

\begin{figure}[t!]
    \centering
    \includegraphics[width=\linewidth]{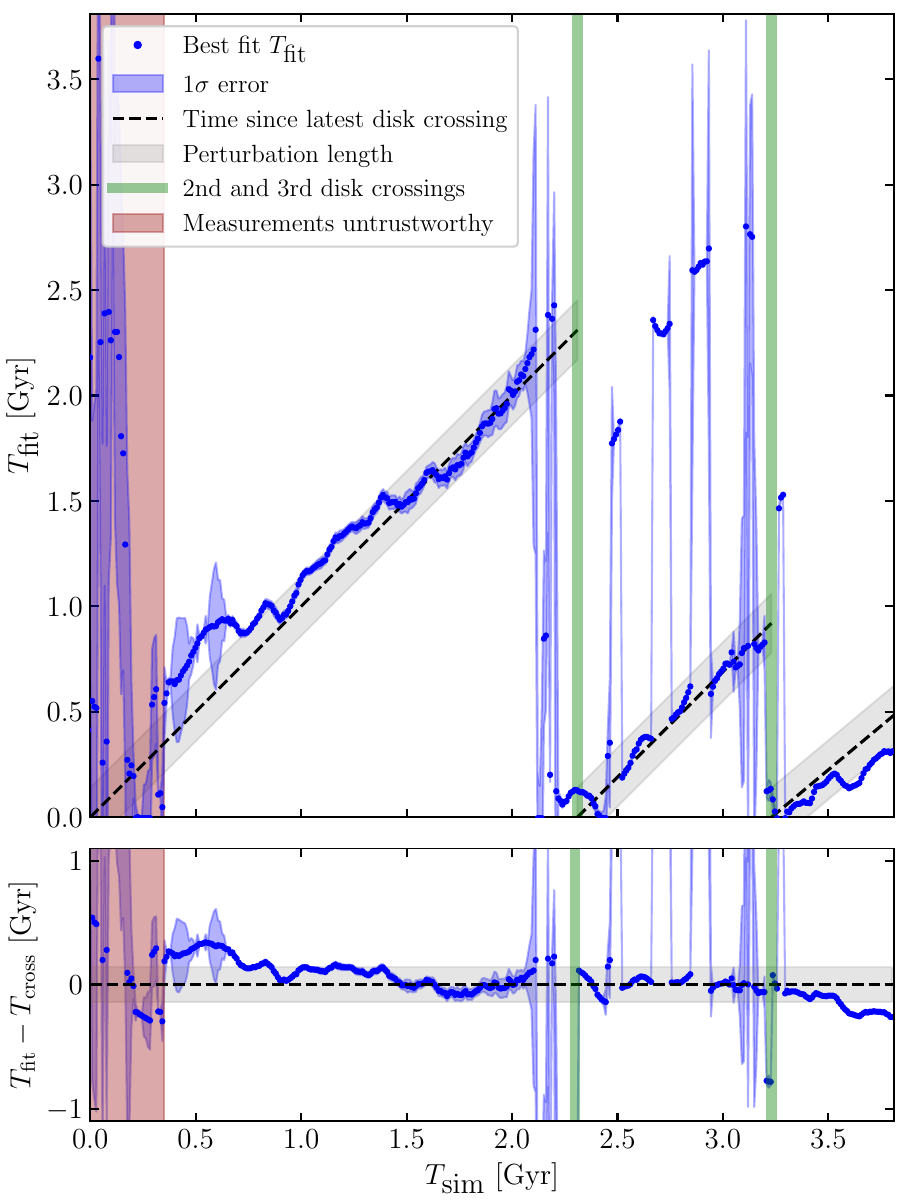}
\caption{\NEW{\textit{Top}: Macro-spiral winding times recovered from each snapshot of the simulation. 
The dashed black lines show the true time since the most recent disk crossing with the gray bands showing a typical perturbation timescale. 
\textit{Bottom}: The residual of the derived $T_{\rm fit}$ values and the most recent disk crossing time. 
The simulation time of 0 \Gyr is defined as the time of the satellite's first disk crossing .
The vertical green bands show the times of the second and third disk crossings within 50 kpc of the Galactic center. 
The brown shaded region at $T_{\rm sim}<0.35 \Gyr$ shows the time span during which our fits are untrustworthy due to the ongoing perturbation.}
}

\label{fig:sim_tfits}
\end{figure}

We analyze the same self-consistent simulation as in Paper I, described in detail therein.
\NEW{At each snapshot (outputted every 9.778 \Myr) we split the disk into 208 bins.} 
We use bin widths of $100 \kpckms$ ($\sim 0.5 \kpc$) for the $J_\phi$ bins and $\pi/8$ for the $\theta_\phi$ bins. 
The innermost bin is centered on \NEW{$J_\phi = 1500 \kpckms$ ($R_g \simeq 6.5 \kpc$)}.
We have \NEW{13 adjacent but non-overlapping $J_\phi$ bins out to $J_\phi = 2700 \kpckms$ ($R_g \simeq 11 \kpc$)}.
\NEW{This is a subset of the bins used in Paper I, where we have cut out some of the outer and inner disk regions that were contributing to inaccurate fits.}

We show the macro-spiral winding results in \figref{sim_tfits}, \NEW{where we define $T_{\rm sim} = 0 \Gyr$ as the time of the first disk crossing.}
\NEW{We also show the true time since the most recent disk crossing as a black dashed line.
The gray band surrounding that line denotes a typical disk crossing timescale, in this case defined by the time the satellite spends within $50 \kpc$ of the galactic center.}
Between the first and second passage, the $T_{\rm fit}$ values largely track the expected values from phase mixing due to differential rotation.
\NEW{Deviations from this trend occur mostly in the few hundred \Myr after the first disk crossing.
This is a period when the satellite is still influencing the disk and notably passes directly above it at $T_{\rm sim} \simeq 0.3 \Gyr$, inducing strong amplitudes in certain disk regions that bias the fit.
We therefore shade in brown the first 0.35 \Gyr of \figref{sim_tfits} and deem these fits as untrustworthy.}

\NEW{From $0.4 \leq T_{\rm sim} \leq 0.7$, $T_{\rm fit}$ is slightly overestimated, although it is sometimes still consistent with the perturbation time.
This overestimate is likely a residual effect of a non-impulsive perturbation.
In fact, for many of the fits in this timespan, it is clear by eye that excess structure from a non-impulsive perturbation is biasing $T_{\rm fit}$ to slightly larger values.
These fits could be tailored individually but we choose to stick with an automated fitting method for the simulation to show the efficacy of a naive application.}

\NEW{After the second and third passages, the figure suggests that each passage creates a new macro-spiral that begins winding. 
The sudden jumps to high $T_{\rm fit}$ values following the second passage are due to the fact that $T_{\rm fit}$ values associated with the first perturbation also fit the data well.}

These results indicate that N-body dynamics do not affect macro-spiral winding.
This is a significant difference from the effect of N-body dynamics on phase-spiral winding discussed in Paper I and provides us with a more accurate way of measuring the time since the perturbation.

However, it is true that this behavior may not extend to much larger perturbations, especially those that induce high-amplitude $z$-oscillations. 
In such cases, stars may experience kicks that significantly change their orbits in a way that affects the in-plane mixing as well, which we do not consider here. 
However, the perturbation induced in this simulation (a $6\times 10^{10} \Msun$ mass satellite crossing the disk at $\simeq 35 \kpc$) is relatively large and suggests that most proposed theories for the phase-spirals' origin would create a macro-spiral that winds up at the rate expected due to differential rotation.

\NEW{Finally, we note that because we apply our method to a simulation with hundreds of timesteps, we require an automated fitting method to derive $T_{\rm fit}$ and $\theta_{\phi,0}$.
When inspecting these fits by eye, we find that $T_{\rm fit}$ is almost always well-recovered when there is a signal with a slope.
However, it is clear from visual inspection that the $\theta_{\phi,0}$ fit is less successful, most likely due to the non-impulsive nature of the perturbation, and could often be improved with a tailored approach to each timestep.
We choose not to do this both because it is impractical and also because the goal of applying this method to the simulation is simply to demonstrate that applying it to observational data will lead to an independent, accurate measurement of the perturbation time.
Since we only have one timestep in the observations, we will be able to tailor the fit to derive both $T_{\rm fit}$ and $\theta_{\phi,0}$ once we have the requisite quality and quantity of data (see \S \ref{sec:observations} and \S \ref{sec:caveats}).
However, because we do not do this for the simulation, we do not show or discuss the derived $\theta_{\phi,0}$ values.}

\subsection{Milky Way macro-spiral results} \label{sec:observations}

In the Milky Way, we do not yet have the ability to characterize phase-spirals across the entire disk.
However, with the release of \Gaia DR3, we are able to detect phase-spirals in a large enough radius around the Sun to apply what we learned from our simulation to observations.
We divide our data sample described in \S \ref{sec:data} into regions with bin widths 0.25 kpc in $R_g$ from 5.5 kpc to 11.25 kpc and 0.1 rad in $\theta_\phi$ from 2.7 rad to 3.6 rad (where the Sun's angular position is defined as $\pi$ rad).

Since we have a limited range of azimuths accessible in the current data, we adapt the method described in \S \ref{sec:tfit_method} to derive $T_{\rm fit}$ from the frequency difference between the location of successive peak amplitudes found as a function of $J_\phi$ at a single azimuth:
\begin{equation} \label{eq:wind_time}
T_{\rm fit} = \frac{\theta_{\phi, \textrm{max}}(J_{\phi1}) - \theta_{\phi, \textrm{max}}(J_{\phi2})}{\Omega_{\phi}(J_{\phi1}) - \Omega_\phi(J_{\phi2})} \equiv \frac{2 \pi}{\Delta\Omega_\phi}.
\end{equation}
Note that we cannot verify that the peaks at a single azimuth are indeed contiguously connected around the disk to form an $m=1$ spiral. This approach instead {\it assumes} that the peaks seen within the data are associated with a single event occurring around one azimuth.

\subsubsection{Results using amplitudes from this work} \label{sec:m1_amp_data}

\begin{figure}[!t]
    \centering
    \includegraphics[width=\linewidth]{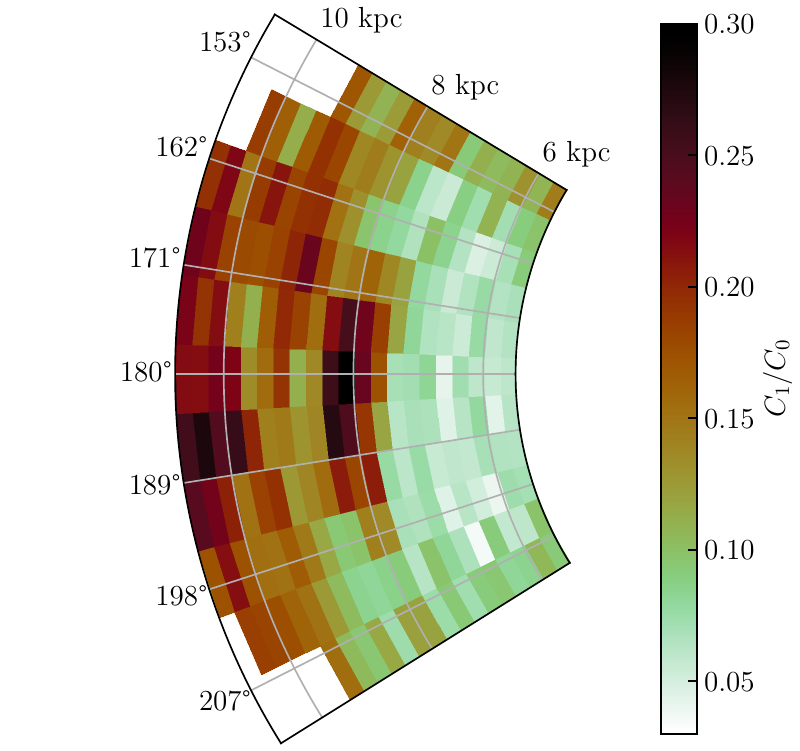}
\caption{The derived one-armed phase-spiral amplitudes in different parts of the Milky Way disk.}
\label{fig:mw_amp}
\end{figure}

We plot the amplitude of the phase-spiral in each region in \figref{mw_amp}.
The results show amplitude generally increasing with guiding radius although there is plenty of variation.
In this limited radial range, it is difficult to conclusively identify macro-spiral ridges.
Nevertheless, there appear to be two amplitude peaks at $R_g\simeq 8.125$ kpc and $R_g\simeq 10.5$ kpc.
This separation implies a macro-spiral winding time of $\simeq 0.9 \Gyr$, assuming the rotation curve given by the  \texttt{MilkyWayPotential2022} potential from the \texttt{Gala} dynamics package \citep{Gala:2017}.

\subsubsection{Results using amplitudes from prior studies} \label{sec:comp}

\begin{figure*}[th!]
    \centering
    \includegraphics[width=\linewidth]{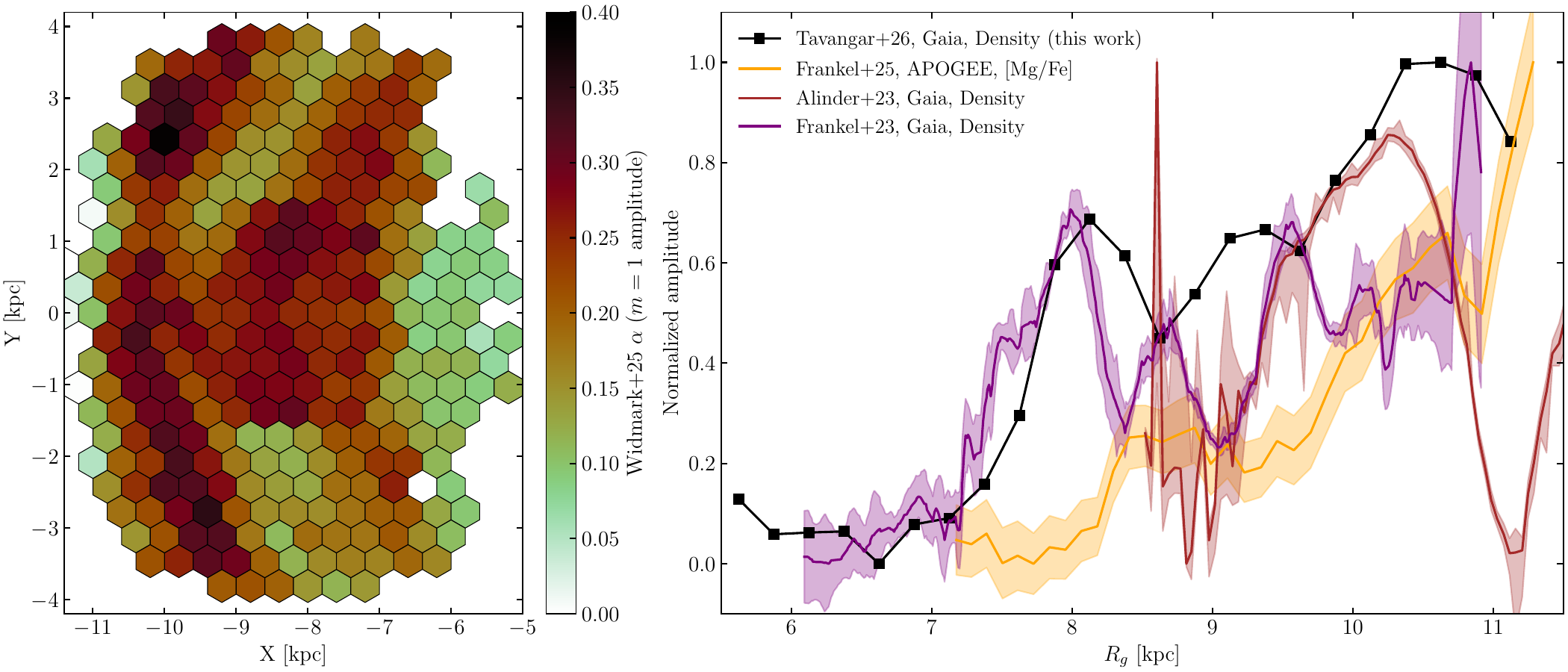}
\caption{Prior works' results on one-armed phase-spiral amplitudes derived from Milky Way data.
\textit{Left:} One-armed phase-spiral amplitudes ($\alpha$) from \citet{Widmark:2025} for many regions near the Sun. 
\textit{Right:} One-armed phase-spiral amplitude as a function of $R_g$ for three previous studies and this work. Different studies use different definitions of amplitude so we normalize all curves to be between 0 and 1. The black squares show the mean one-armed amplitude from this work for a given $R_g$ bin.
In all other works, amplitudes are derived in $L_z$ bins, which we convert to $R_g$ using the rotation curve of the Milky Way associated with the \texttt{MilkyWayPotential2022} in \texttt{Gala} \citep{Gala:2017}.}
\label{fig:amp_comp}
\end{figure*}

We now compare our amplitude results to prior studies and extract macro-spiral winding times from their measurements.
The only previous study that has published phase-spiral amplitudes for 2D bins across the disk is \citet{Widmark:2025}. 
We show their amplitude results in the left panel of \figref{amp_comp}, although we caution, as they do, that amplitude is highly affected by systematics such as selection effects.
They also choose to work in physical space rather than angle-action space, which does not affect the macro-spiral winding derivation but does mean that they use a proper motion sample with estimated rather than measured radial velocities beyond 1.6 kpc.
We immediately notice two arcing bands of higher amplitude phase-spirals, precisely the pattern we expect the macro-spiral to resemble.
By eye the amplitudes appear to peak at $R\simeq 7.5$ and $R\simeq 10$, 
indicating a perturbation time $\simeq 0.8 \Gyr$ ago.

Other studies have plotted the amplitude of the Milky Way phase-spirals as a one-dimensional function of angular momentum \citep{Alinder:2023, Frankel:2023, Frankel:2025}.
We show the (normalized) phase-spiral amplitudes as a function of $R_g$ for these studies in \figref{amp_comp}.
In two of the three studies \citep{Alinder:2023, Frankel:2025} it is difficult to identify two amplitude peaks with which to calculate the winding time.
In the third study \citep{Frankel:2023}, there exist two prominent peaks at $R_g \simeq 8 \textrm{ and } 9.55$ kpc, consistent with a perturbation time $\simeq 1.25$ \Gyr ago.

Finally, a couple previous studies have used macro-spirals in other parameters (\ie unrelated to the phase-spiral) to derive disk perturbation times using a similar method.
Notably, both \citet{Antoja:2022} and \citet{Lambert:2026} recover a perturbation $\simeq 1 \Gyr$ ago\footnote{\citet{Antoja:2022} find a very broad a range for this perturbation time of $0.8-2.1 \Gyr$ ago}, consistent with our results.

\subsection{Implications for individual phase-spiral delays} \label{sec:delays}

We showed in Paper I that phase-spirals in self-consistent simulations are less tightly wound than would be expected from pure phase mixing theory, and that the delays in their winding vary as a function of guiding radius.
However, the details of the dynamics involved in generating these delays remain underexplored.
One significant benefit of having an independent and robust measurement of the phase-spiral perturbation time is that it allows us to know the delay in individual phase-spirals across large parts of the Milky Way.
For example, combining our $\simeq 0.9 \Gyr$ estimate for the macro-spiral winding time with the phase-spiral winding time measurements from \citet{Widmark:2025} (shown in a 2D plot in Figure 10 of Paper I) suggests that phase-spiral winding is delayed by up to $700 \Myr$ in the inner disk ($6 \lesssim R \lesssim 8 \kpc$) but is negligibly delayed in the outer disk ($R_g \gtrsim 10 \kpc$).
Such estimates will be invaluable for improving our understanding of both the dynamics of N-body interactions and the Milky Way disk potential.

\subsection{Caveats} \label{sec:caveats}

We caution that these estimates for the macro-spiral winding time should not be interpreted as robust measurements for two main reasons.
First, the region of the disk with appropriate \Gaia data for these studies is too limited to observe more than two peaks in $R_g$ or for researchers to be sure that these peaks belong to a single continuous macro-spiral. 
\Gaia DR4 will provide radial velocities for many more stars, with which we will be able to measure phase-spirals across nearly the entire Milky Way disk.
Second, no study has yet rigorously accounted for the dust extinction and other selection effects in a detailed way that also allows us to map a larger region.
For instance, this study, along with \citet{Frankel:2023}, primarily chooses to account for the selection function by making a distance cut to reduce the effect of dust extinction.
Clearly, this strategy is not conducive to analyzing a larger part of the disk, meaning that future studies will need to address this shortcoming.

\section{Conclusions} \label{sec:conclusion}

\subsection{Summary of theoretical explorations so far.}
In this series of papers, we use a fully self-consistent simulation of a perturbed galactic disk to explore what large-scale coherence in phase-spiral morphology can reveal about the recent history of that galaxy and its potential.
In Paper I we found delays in the onset of phase-spiral winding and that the winding rate varied systematically with radius.

This paper focuses on the patterns formed when phase-spiral {\it amplitudes} are mapped in radial and azimuthal bins across a galactic disk.
We demonstrate that the macroscopic spiral formed from connecting the maxima of phase-spiral amplitudes following a perturbation winds up at a rate following the differential rotation of the disk.

\subsection{Future prospects: in theory}
There is currently no consensus about the observed phase-spirals' origins, including i) what event (or events) seeded these structures and ii) when did this event (or these events) happen. Our conclusions so far suggest that the robust measurements of phase-spiral amplitude across the disk is a way forward. 
In the most optimistic terms:
\begin{itemize}
    \item the unwinding time ($T_{\rm fit}$ for any macro-spiral in amplitudes (i.e. local maxima whose azimuth vary smoothly with radius) gives an estimate of the time since perturbation;
    \item unwinding the macro-spiral can indicate the azimuthal location of the perturbing event ($\theta_{\phi,0}$);
    \item the variation of amplitude with radius can reveal the galactocentric radial location of the perturbing event;
    \item the offset between the unwinding times derived from the micro-spirals relative to the macro-spiral provides a constraint on the nature of the local galactic disk.
\end{itemize}

\subsection{Future prospects: in practice}
While it is interesting to think about the implications of our analysis of simulations in theory, the reality is more complicated. 
For the moment, phase-spiral amplitude measurements are inconsistent across different analyses due to insufficient data and differences in methods.
For example, we recover a range of perturbation times between 800 \Myr and 1.25 \Gyr ago --- implying phase-spiral winding delays of 600--1000 \Myr in the inner disk --- depending on which phase spiral amplitude measurements we used.

Currently, the missing ingredient is the ability to make robust measurements of phase-spiral properties over large regions of the Milky Way disk.
Luckily, the prospects are bright. 
Future \Gaia data releases will allow us to map individual phase-spiral properties over much larger areas of the disk and we may discover one or many ``macro-spiral'' groups, formed from sets of coherently varying phase-spirals.

\vspace{24pt}

% \textbf{Future work}
% \item Phase angle analysis (see Widmark+25)???
% \item There are two-armed spirals in the n-body simulation, particularly in the outer disk. Do they give us any interesting info?

% \KT{Fit for the shape of the rotation curve if we have more peaks or an independent measurement of the perturbation time? (Where) should this go in?}

%% Also note that the akcnowlodgment environment does not support long amounts of text. If you have a lot of people and institutions to acknowledge, do not use this command. Instead, create a new \section{Acknowledgments}.
\begin{acknowledgments}
The authors thank the referee for useful comments that greatly improved the paper.
KT thanks Neige Frankel, Simon Alinder, and Paul McMillan for providing the data from their previous works to help create \figref{amp_comp}.
We thank the EXP collaboration and in particular Martin Weinberg, Mike Petersen, and Chris Hamilton for useful discussions about the series of papers overall. We also thank the Nearby Universe group within the Center for Computational Astrophysics (CCA) at the Flatiron Institute, and the participants of the \textit{Winding, Unwinding, and Rewinding the \Gaia phase-spiral} workshop in August 2025 for useful discussions. We also thank the computing resources at the CCA for running and hosting the simulations used in this work. 
K.V.J. is supported by Simons Foundation grant 1018465.
JH acknowledges the support of a UKRI Ernest Rutherford Fellowship ST/Z510245/1.
AW is supported by the European Union’s Horizon 2020 research and innovation program, under the Marie Sklodowska-Curie grant agreement number 101106028. 
VK acknowledges the support of UKRI studentship ukri1785.
\end{acknowledgments}

%\software{astropy \citep{2013A&A...558A..33A,2018AJ....156..123A},}

%% pdflatex main.tex
%% bibtext main
%% pdflatex main.tex
%% pdflatex main.tex

\bibliography{main}{}
\bibliographystyle{aasjournal}

% \appendix

\end{document}